\documentclass[11pt,fleqn]{article}

\usepackage{amsmath,amssymb,amsthm,cite,graphics,epsfig,enumerate}

\newcommand{\p}{\partial}

\newcommand{\todo}[1][\null]{\ensuremath{\clubsuit}}
\newcommand{\checked}[1][\null]{\ensuremath{\diamond}}
\newcommand{\noprint}[1]{}

\newcounter{tbn}

\newcounter{mcasenum}

\newtheorem{theorem}{Theorem}

\newtheorem{corollary}{Corollary}
\newtheorem{proposition}{Proposition}
\newtheorem*{proposition*}{Proposition}
{\theoremstyle{definition}

\newtheorem{example}{Example}
\newtheorem{remark}{Remark}

}

\begin{document}\allowdisplaybreaks
\begin{center}
\par\noindent {\LARGE\bf
Group classification of variable coefficient\\ generalized Kawahara  equations
\par}

{\vspace{5mm}\par\noindent\large Oksana Kuriksha$^{\dag }$, Severin Po{\v s}ta$^{\ddag }$ and Olena~Vaneeva$^{\S }$\footnote{Corresponding author.}
\par\vspace{1mm}\par}
\end{center}

{\par\noindent\it\small
${}^\dag$\ Petro Mohyla Black Sea State University, 10, 68 Desantnykiv Street, 54003 Mykolaiv, Ukraine\\[1ex]
${}^\ddag$\ Department of Mathematics, Faculty of Nuclear Sciences and Physical Engineering,\\[1ex]
 $\phantom{{}^\ddag}$\ Czech Technical University in Prague, 13 Trojanova Str., 120 00 Prague, Czech Republic \\[1ex]
${}^\S$\ Institute of Mathematics of the National Academy of Sciences of Ukraine,\\[1ex]
$\phantom{{}^\S}$\ 3 Tereshchenkivska Str., 01601 Kyiv-4, Ukraine
}

{\vspace{2mm}\par\noindent
$\phantom{{}^\dag{}\;}\ $E-mails: \it oksana.kuriksha@gmail.com, severin.posta@fjfi.cvut.cz,
vaneeva@imath.kiev.ua
\par}

{\vspace{5mm}\par\noindent\hspace*{5mm}\parbox{150mm}{\small
An exhaustive group classification of variable coefficient generalized Kawahara equations is carried out. As a result, we derive new variable coefficient nonlinear models admitting Lie symmetry extensions.
All inequivalent Lie reductions of these equations to ordinary differential equations are performed. We also present some examples on the construction of exact and numerical solutions.
}\vspace{4mm}\par}

\section{Introduction}
In this paper we study generalized Kawahara equations with time-dependent coefficients
\begin{gather}\label{eq_ggKawahara}
u_t+\alpha(t)u^nu_x+\beta(t)u_{xxx}+\sigma(t)u_{xxxxx}=0
\end{gather}
from the Lie symmetry point of view. Here $n$ is an arbitrary nonzero integer, $\alpha$, $\beta$ and $\sigma$ are smooth nonvanishing functions of the variable $t$.

It is widely known that Lie (point) symmetries of differential equations (DEs) give a powerful tool for finding exact solutions and this is one of the most successful applications of geometrical studies of DEs~\cite{Kamran,Olver1986,Ovsiannikov1982}. At the same time it is much less known
that  Lie symmetries can be served also as a selection principle for equations which are important for applications among wide set of possible models.
All fundamental equations of mathematical physics, e.g., the
 Maxwell, Schr\"odinger, Newton, Laplace, Euler--Lagrange, d'Alembert, Lam\'{e}, Hamilton--Jacobi equations, etc., have
nontrivial symmetry properties, i.e., they admit multi-dimensional Lie invariance algebras. Moreover,  many equations of mathematical physics can be derived just from requirement of invariance with respect to a transformation group. For example, there is only one system of Poincar\'e-invariant
first-order partial differential equations  for two real vectors $\mathbf{{E}}$
and $\mathbf{{H}}$, and this is the system of Maxwell equations~\cite{FN}.  The important problem of classification of all possible Lie symmetry extensions for
equations from a given class with respect to the equivalence
group of the class is called the group classification problem~\cite{Ibr,Ovsiannikov1982,popo2004b}.

If $\alpha$, $\beta$ and $\sigma$ are not functions but constants, equations~\eqref{eq_ggKawahara} become classical models appearing in the solitary waves theory. Here we present a brief overview on applications of Kawahara equations and the related results.
In the usual sense, solitary waves are nonlinear waves of constant form which decay rapidly in their tail regions. The rate of this decay is usually exponential. However, under critical conditions in dispersive systems (e.g.,  the magneto-acoustic waves in plasmas, the waves with surface tension, etc.), unexpected rise of weakly nonlocal solitary waves occurs. These waves consist of a central core which is similar to that of classical solitary waves, but they are accompanied by copropagating oscillatory tails which extend indefinitely far from the core with a nonzero constant amplitude. In order to describe and clarify the properties of these waves Kawahara introduced generalized nonlinear dispersive equations which have a form of the KdV equation with an additional fifth order derivative term, namely,
\[
u_t+\alpha u u_x +\beta u_{xxx}+\sigma u_{xxxxx}=0,
\]
where  $\alpha$, $\beta$ and $\sigma$ are nonzero constants~\cite{Hasimoto,Kawahara1972}.  This equation was heavily studied from different points of view. The exact solitary wave solution was presented in~\cite{Yamamoto&Takizawa1981}. In \cite{Hunter1988} the existence of travelling wave solutions of the Kawahara equation being considered as a formal asymptotic approximation for water waves with surface tension was shown.
In \cite{Boyd1991} various numerical computations of both infinite interval and spatially periodic solutions to  a  one-dimensional  wave  equation  which models  capillary-gravity  waves  were done. Using techniques of exponential asymptotics it was shown in \cite{grimshaw&joshi1995} that solitary wave solutions of the Kawahara equation form a one-parameter family characterized by the phase shift of the trailing oscillations. An explicit
asymptotic formula relating the oscillation amplitude to the phase shift was obtained therein. Solvability of the Cauchy problem (local and global existence) of the Kawahara equation was studied in~\cite{cui,Iguchi}. Various studies on behavior of solutions of the Kawahara equations were presented, e.g.,
in~\cite{Bakholdin2001,Bakholdin2013,Djidjeli1995,Dubrovin2011,Gunney&Li&Olver1999,Yamamoto1989}.

Generalized constant coefficient models related to the Kawahara equation have appeared later. For example,  long waves in a shallow liquid under ice cover in the presence of tension or
compression were described by the equation $u_t+u_x+\alpha u u_x +\beta u_{xxx}+\sigma u_{xxxxx}=0$~\cite{Marchenko1988,Tkachenko&Yakovlev1999}. This equation is similar to the classical Kawahara equation with respect to the simple changes of variables: $\tilde x=x-t,$ where $t$ and $u$ are not transformed, or $\tilde u=1+\alpha u$, where $t$ and $x$ are not transformed.
An analytical theory of radiating and stationary solitons, satisfying the modified Kawahara equations
\[
u_t+\alpha u^n u_x +\beta u_{xxx}+\sigma u_{xxxxx}=0,
\]
 where $\alpha$, $\beta$ and $\sigma$ are nonzero constants, and $n\in\mathbb{N},$ was given in~\cite{Karpman1994}. The stability in the sense of Lyapunov for solitons described by these equations  was studied in~\cite{Karpman1996}. 

 We note that neither the classical Kawahara equation nor its generalization adduced above are integrable by the inverse scattering
transform method~\cite{MikhailovShabatSokolov1991,encycl}.

Last time much attention is paid to  variable coefficient models, like variable coefficient KdV, Burgers, and Schr\"odinger equations~\cite{Popovych&Vaneeva2010}. This is due to the fact that variable coefficient equations can model certain real-world phenomena with more accuracy than their constant coefficient counterparts.
In the recent paper~\cite{Kaur&Gupta} Lie symmetries were applied for finding exact solutions of variable coefficient Kawahara and modified Kawahara equations, which are of the form~\eqref{eq_ggKawahara} with $n=1$ and $n=2$,
respectively. The presence of three arbitrary coefficients depending on~$t$ makes the task of finding Lie symmetries too difficult to get complete results without reducing  the number of variable coefficients by equivalence transformations. That is why only few results on Lie symmetries were derived in~\cite{Kaur&Gupta}.
In the present paper we show that the use of such transformations is a cornerstone in the complete solution of the problem.

The structure of this paper is as follows. All point transformations between equations from class~\eqref{eq_ggKawahara} (so-called admissible transformations) are exhaustively described in Section~2. Possibilities of reducing equations from this class to a simpler form are discussed therein.
We choose the gauge $\alpha=1$ and justify that it is optimal. The classical algorithm based on applying the Lie invariance criterion to~\eqref{eq_ggKawahara} and subsequent study of compatibility and direct integration of the derived determining equations is utilized in Section~3 to get the complete group classification. As a result, new variable coefficient models with nontrivial Lie symmetry properties are singled out from~\eqref{eq_ggKawahara}.
Section~4 is devoted to the classification of Lie reductions and finding exact and numerical solutions for the variable coefficient generalized Kawahara equations. We give some final remarks and discuss problems for further investigation in Section~5.

\section{Admissible transformations}
If two differential equations  are connected by a change of variables (a point transformation),  they are called similar equations~\cite{Ovsiannikov1982}.
Then related objects like, e.g., exact solutions, conservation laws, different kinds of symmetries of such equations,  are also similar. If they are known
for one of these equations, then  their counterparts for the other equation can be derived using the aforementioned transformation.
This is why when one deals with a  class of differential equations parameterized by arbitrary elements (constants or functions),
it is highly important to study relations between fixed equations from this class that are induced by point transformations.
Such similarity relations are called  in the literature allowed~\cite{Winternitz92},  form-preserving~\cite{Kingston&Sophocleous1998}, and  admissible~\cite{popo2010a} transformations.
An admissible transformation can be interpreted as a triple consisting of two fixed equations from a class
and a~point transformation that links these equations.
The set of admissible  transformations considered with the standard operation of composition of  transformations is also called the equivalence groupoid~\cite{Popovych&Bihlo2012}.

Equivalence transformations generate a subset in  a set of admissible transformations.    It is important that admissible transformations are not necessarily related to a group structure, but equivalence transformations always form a group. An equivalence transformation  applied to any equation from the class always maps it to another equation from the same class. In other words, equivalence transformations preserve differential structure of the class. At the same time, an admissible transformation may exist only for a specific pair of equations from the class under consideration. For example, the point transformation $t'=e^{bt}/b$, $x'=x,$ $u'=u-bt$ links equations $u_t=(e^u)_{xx}+ae^u+b$ and ${u'}_{t'}=(e^{u'})_{x'x'}+ae^{u'}$, where $a$ and $b$ are arbitrary constants with $b\neq0$~\cite{Ibr}. Both these equations are members of the class
$\mathcal L\colon\ u_t=(e^u)_{xx}+Q(u)$, where $Q$ is a smooth function of $u$. Acting on other equation from this class, e.g., on $u_t=(e^u)_{xx}+e^{2u}+b$, this transformation maps it to the equation ${u'}_{t'}=(e^{u'})_{x'x'}+b\,t'e^{2u'}$, that is not constant coefficient one and does not belong to the class~$\mathcal L$.

By Ovsiannikov, the equivalence group consists of the nondegenerate point transformations
of the independent and dependent variables and of the arbitrary elements of the class,
where transformations for independent and dependent variables are projectible on the space of these variables~\cite{Ovsiannikov1982}. After appearance of other kinds of equivalence group the one used by Ovsiannikov is called now usual equivalence group.
If the transformations for independent and dependent variables involve  arbitrary
elements, then the corresponding equivalence group is called the generalized equivalence group~\cite{Meleshko1994}.
If new arbitrary elements appear to depend on old ones in a nonlocal way (e.g., new arbitrary elements are expressed via integrals of old ones),
then the corresponding equivalence group is called extended~\cite{mogran}. Generalized extended equivalence group possesses both the aforementioned properties.
A~number of examples of usage of different kinds of equivalence groups are presented, e.g., in~\cite{ips2010a,vane2012b}.

If any admissible transformation in a given class is induced by a transformation from its equivalence group
(usual / generalized / extended / generalized extended), then this class is called normalized in the corresponding sense.

We search for admissible transformations in class~\eqref{eq_ggKawahara} using the direct method~\cite{Kingston&Sophocleous1998}.
Suppose that equation~\eqref{eq_ggKawahara} is similar  to an equation from the same class,
\begin{equation}\label{eq_ggKawahara_tilda}
\tilde u_{\tilde t}+\tilde\alpha(\tilde t) \tilde u^{\tilde n}\tilde u_{\tilde x}+\tilde\beta(\tilde t) \tilde u_{\tilde x\tilde x\tilde x}+\tilde\sigma(\tilde t) \tilde u_{\tilde x\tilde x\tilde x\tilde x\tilde x}=0,
\end{equation}
with respect to a nondegenerate point~transformation in the space of variables $(t,x,u)$.
 We can restrict ourselves by consideration of point transformations of the form
\begin{gather}\label{EqEquivtransOfvcKdVlikeSuperclass}
\tilde t=T(t),\quad \tilde x=X^1(t)x+X^0(t),\quad \tilde u=U^1(t,x)u+U^0(t,x),
\end{gather}
where $T$, $X^1$, $X^0$, $U^1$ and $U^0$ are arbitrary smooth functions of their variables with $T_tX^1U^1\neq0$.
This restriction does not lead to any loss of generality for a subclass of the normalized class of evolution equations,
\begin{gather*}
u_t=F(t)u_n+G(t,x,u,u_1,\dots,u_{n-1}),\quad F\ne0,\quad G_{u_iu_{n-1}}=0,\ i=1,\dots,n-1,
\end{gather*}
where $n\geqslant 2$,
$
u_{n}=\frac{\p^n u}{\p x^n},
$
$F$ and $G$ are arbitrary smooth functions of their variables~\cite{VPS2013}.
Under transformations~\eqref{EqEquivtransOfvcKdVlikeSuperclass} partial derivatives involved
in~\eqref{eq_ggKawahara} are transformed as follows
\begin{gather*}\arraycolsep=0ex
\begin{array}{l}
\tilde u_{\tilde t}\!=\!\dfrac{1}{T_t}\left(U^1_tu+U^1u_t+U^0_t\right)-\dfrac{X^1_tx+X^0_t}{T_tX^1}\left(U^1_xu+U^1u_x+U^0_x\right),\\[2ex]
\tilde u_{\tilde x}\!=\!\dfrac{U^1_xu+U^1u_x+U^0_x}{X^1},\qquad
\tilde u_{\tilde x\tilde x\tilde x}\!=\!\dfrac{U^1_{xxx}u+3U^1_{xx}u_x+3U^1_{x}u_{xx}+U^1u_{xxx}+U^0_{xxx}}{(X^1)^3},\\[2ex]
\tilde u_{\tilde x\tilde x\tilde x\tilde x\tilde x}\!=\!\dfrac{U^1_{xxxxx}u+5U^1_{xxxx}u_x+10U^1_{xxx}u_{xx}+10U^1_{xx}u_{xxx}+5U^1_xu_{xxxx}+U^1u_{xxxxx}+U^0_{xxxxx}}{(X^1)^5},
\end{array}
\end{gather*}
Rewriting~\eqref{eq_ggKawahara_tilda} in terms of the untilded variables,
we further substitute $u_t=-\alpha(t)u^nu_x-\beta(t)u_{xxx}-\sigma(t)u_{xxxxx}$ to the obtained equation
in order to confine it to the manifold defined by~\eqref{eq_ggKawahara} in the fifth-order jet space
with the independent variables $(t,x)$ and the dependent variable~$u$.
Splitting the obtained identity with respect to the derivatives of $u$ leads to the determining
equations on the functions~$T$, $X^1$, $X^0$, $U^1$ and $U^0$.
In particular, we get the following conditions \[\tilde n=n,\quad U^1_x=0,\quad \tilde\beta T_t-\beta(X^1)^3=0,\quad \tilde\sigma T_t-\sigma(X^1)^5=0.\]
The further splitting depends on whether $n\neq1$ or $n=1$. The rest of the determining equations in these cases are
the following
\begin{gather*}
\boldsymbol{n\neq1}\colon\quad U^0=U^1_t=X^1_t=X^0_t=0, \quad \tilde\alpha\left(U^1\right)^nT_t-\alpha X^1=0;\\
\boldsymbol{n=1}\colon\quad
U^1_tX^1+\tilde\alpha T_tU^1U^0_x=0, \quad \tilde\alpha U^1T_t-\alpha X^1=0, \quad
\tilde\alpha T_t U^0-X^1_tx-X^0_t=0, \\ U^0_t\left(X^1\right)^5-\left(X^1_tx+X^0_t\right)U^0_x\left(X^1\right)^4+\tilde\alpha T_t U^0U^0_x\left(X^1\right)^4+\tilde\beta T_t U^0_{xxx}\left(X^1\right)^2+\tilde\sigma T_t U^0_{xxxxx}=0.
\end{gather*}
Solving these equations we get exactly the statements presented in Theorems 1 and 2.

\begin{theorem}The usual equivalence group~$G^{\sim}$ of class~\eqref{eq_ggKawahara} consists of the transformations
\begin{gather*}
\tilde t=T(t),\quad \tilde x=\delta_1x+\delta_2,\quad
\tilde u=\delta_3 u, \\[1ex]
\tilde \alpha(\tilde t)=\frac{\delta_1}{{\delta_3}^{\!n}
T_t}\alpha(t), \quad\tilde\beta(\tilde
t)=\dfrac{\delta_1^3}{T_t}\beta(t),\quad \tilde\sigma(\tilde
t)=\dfrac{\delta_1^5}{T_t}\sigma(t),\quad\tilde n=n,
\end{gather*}
where  $\delta_j,$ $j=1,2,3,$ are arbitrary constants with
$\delta_1\delta_3\not=0$, $T$ is an arbitrary smooth function with $T_t\neq0.$
\end{theorem}

\begin{theorem} The generalized extended equivalence group~$\hat G^{\sim}_{n=1}$ of the class
\begin{equation}\label{eq_Kawahara_n=1}
u_t+\alpha(t)uu_x+\beta(t)u_{xxx}+\sigma(t)u_{xxxxx}=0
\end{equation} is formed by the transformations
\begin{gather*}
\tilde t=T(t),\quad \tilde x=(x+\delta_1)X^1+\delta_0,\quad
\tilde u=\frac{\delta_2}{X^1} u -\delta_2\delta_3(x+\delta_1), \\[1ex]
\tilde \alpha(\tilde t)=\frac{(X^1)^2}{\delta_2T_t}\alpha(t),\quad\tilde\beta(\tilde t)=\dfrac{(X^1)^3}{T_t}\beta(t),\quad\tilde\sigma(\tilde t)=\dfrac{(X^1)^5}{T_t}\sigma(t),
\end{gather*}
where $X^1=(\delta_3\int \alpha(t) {\rm d}t+\delta_4)^{-1},$ $\delta_j,$ $j=0,\dots,4,$ are arbitrary constants with
$\delta_2(\delta_3{}^2+\delta_4{}^2)\not=0$; $T=T(t)$ is a smooth function with $T_t\neq0$. The usual equivalence group~$G^{\sim}_{n=1}$ of class~\eqref{eq_Kawahara_n=1} comprises the above transformations with $\delta_1=\delta_3=0.$
\end{theorem}
\begin{theorem}
A variable coefficient equation from class~\eqref{eq_ggKawahara} is reducible to constant coefficient equation from the
same class if and only if the coefficients $\alpha$, $\beta$ and $\sigma$ satisfy the conditions
\begin{gather}\label{criterion1}
\left(\frac\beta\alpha\right)_t=\left(\frac{\sigma\vphantom{\beta}}\alpha\right)_t=0, \quad\mbox{for}\quad n\ne1,\\[1.5ex]
\label{criterion2}
\left(\frac1{\alpha}\left(\frac\beta\alpha\right)_t\right)_t=0,\quad
\left(\frac{\sigma\vphantom{\beta}\alpha^2}{\beta^3}\right)_t=0,\quad\mbox{for}\quad n=1.
\end{gather}
\end{theorem}

The presence of the arbitrary function $T(t)$ in the equivalence
transformations adduced in Theorems 1 and 2 allows one to gauge
one of the arbitrary functions $\alpha$, $\beta$ and $\sigma$ to
a simple constant value, e.g., to 1.
An interesting question is which one of the three possible gauges is preferable for further consideration.
Class~\eqref{eq_Kawahara_n=1} with $\beta=1$ or $\sigma=1$ is still normalized only in the generalized extended sense, since transformations of independent and dependent variables still involve $\int \alpha(t){\rm d}t$.
At the same time class~\eqref{eq_Kawahara_n=1} with $\alpha=1$ is normalized with respect to its usual equivalence group, as $X^1$ appearing in Theorem~2 in this case takes the form
$X^1=(\delta_3t+\delta_4)^{-1}.$
This is why we can expect that in the case $n=1$ it is easier to carry out the group classification
under the gauge $\alpha=1$ rather than under other possible gauges. If $n\neq1$ all the three suggested gauges look equally convenient, and we choose the gauge $\alpha=1$ just to present the group classification in the uniform way.

The gauge $\alpha=1$  is realized by the point
transformation
\begin{equation}\label{eq_gauge}
\hat t=\int\!\alpha(t)\,{\rm d}t, \quad \hat x=x,\quad\hat u=u.
\end{equation} Then
class~\eqref{eq_ggKawahara} is mapped to its subclass  with $\hat
\alpha=1$, $\hat\beta=\beta/\alpha$ and $\hat\sigma=\sigma/\alpha$.
Therefore, without loss of generality we can restrict ourselves to the study of the class
\begin{gather}\label{eq_gKawahara}
u_t+u^nu_x+\beta(t)u_{xxx}+\sigma(t)u_{xxxxx}=0,
\end{gather}
since all results on symmetries, conservation laws, classical solutions and other related objects for equations~\eqref{eq_ggKawahara} can be found
using the similar results derived for equations from class~\eqref{eq_gKawahara}.

To derive the equivalence group for subclass of class~\eqref{eq_ggKawahara} with $\alpha=1$ we set $\tilde\alpha=\alpha=1$ in the transformations presented in Theorems~1 and~2.
\begin{corollary}The generalized equivalence group~$\hat G^{\sim}_{\alpha=1}$ of class~\eqref{eq_gKawahara} comprises  the transformations
\begin{gather}\label{tr_equiv_cor2}
\begin{array}{@{}l}
\tilde t=\delta_1\delta_3{}^{-n}t+\delta_0,\quad \tilde x=\delta_1x+\delta_2,\quad
\tilde u=\delta_3 u, \\[1ex]
\tilde\beta(\tilde t)=\delta_1{}^2\delta_3{}^{n}\beta(t),\quad\tilde\sigma(\tilde t)=\delta_1{}^4\delta_3{}^n\sigma(t),\quad\tilde n=n,
\end{array}
\end{gather}
where  $\delta_j,$ $j=0,1,2,3,$ are arbitrary constants with
$\delta_1\delta_3\not=0$.
\end{corollary}
\begin{remark} If we assume that the constant~$n$  varies in class~\eqref{eq_gKawahara}, then the equivalence group~$\hat G^{\sim}_{\alpha=1}$  is 
generalized since $n$ is involved explicitly in the  transformation of the variable $t$.
From the other hand, $n$ is invariant under the action of transformations from the equivalence group,
so class~\eqref{eq_gKawahara} can be considered as the union of all its subclasses with fixed~$n$.
For each such subclass the group $\hat G^{\sim}_{\alpha=1}$ is
usual equivalence group.
\end{remark}
In the case $n=1$ we put $\alpha=\tilde\alpha=1$ in transformation from Theorem~2 and redenote the constants $\delta_j$, $j=0,\dots,4,$ to write the transformations in a more compact form.

\begin{corollary}
The usual equivalence group~$G^\sim_{\alpha=n=1}$ of the class
\begin{gather}\label{eq_gKawahara_n1}
u_t+uu_x+\beta(t)u_{xxx}+\sigma(t)u_{xxxxx}=0
\end{gather}
 consists of the transformations
\begin{gather}\label{EqZhang2007ReducedEquivGroup}\arraycolsep=0ex
\begin{array}{l}
\tilde t=\dfrac{at+b}{ct+d},\qquad
\tilde x=\dfrac{e_2x+e_1t+e_0}{ct+d},\qquad
\tilde u=\dfrac{e_2(ct+d)u-e_2cx-e_0c+e_1d}\Delta,\\[2ex]
\tilde \beta=\dfrac{e_2{}^3}{ct+d}\dfrac {\beta}\Delta,\qquad\tilde \sigma=\dfrac{e_2{}^5}{(ct+d)^3}\dfrac {\sigma}\Delta,
\end{array}
\end{gather}
where $a$, $b$, $c$, $d$, $e_0$, $e_1$ and $e_2$ are arbitrary constants with $\Delta=ad-bc\ne0$ and $e_2\ne0$,
the tuple  $(a,b,c,d,e_0,e_1,e_2)$ is defined up to a nonzero multiplier
and hence without loss of generality we can assume that $\Delta=\pm1$.
\end{corollary}

\noprint{Equations from class~\eqref{eq_gKawahara_n1}  are similar only if they are $G^\sim_{\alpha=n=1}$-equivalent.
Moreover, all admissible transformations in this class are generated by transformations from $G^\sim_{\alpha=n=1}$,
i.e., the class~\eqref{eq_gKawahara_n1} is normalized in the usual sense.
(The initial class~\eqref{eq_Kawahara} is normalized only with respect to the extended generalized equivalence group
and class~\eqref{eq_ggKawahara} possesses no normalization properties.)
This implies the following claim:
\emph{An equation of form~\eqref{eq_gKawahara_n1} is similar to the Kawahara equation if and only if $\beta=c_1t+c_0$, and $\sigma=(c_1t+c_0)^3$.
Any transformation realizing the similarity belongs to~$G^\sim_{\alpha=n=1}$.}
Therefore, an equation of form~\eqref{eq_Kawahara} is reduced to the Kawahara equation by a point transformation if and only if
\begin{equation}\label{EqCondOnEquiv}
\beta(t)=\alpha(t)\left(c_1\int \alpha(t)\,dt+c_0\right),\quad \sigma(t)=\alpha(t)\left(c_1\int \alpha(t)\,dt+c_0\right)^3,
\end{equation}
where $c_0$ and $c_1$ are constants, $(c_0,c_1)\ne(0,0)$.}

\section{Lie symmetries}
The group classification of equations of the form~\eqref{eq_gKawahara} with $n\neq1$ up to $\hat G^{\sim}_{\alpha=1}$-equivalence (resp. up to $G^{\sim}_{\alpha=n=1}$-equivalence if $n=1$) coincides with
the group classification of equations of the form~\eqref{eq_ggKawahara} with $n\neq1$ up to $G^{\sim}$-equivalence (resp. up to $\hat G^{\sim}_{n=1}$-equivalence if $n=1$).
In order to carry out the group classification of class~\eqref{eq_gKawahara}
we use the  classical algorithm based on direct integration of determining equations implied by the infinitesimal invariance criterion~\cite{Olver1986,Ovsiannikov1982} (see modern discussion on algebraic method of group classification, e.g., in~\cite{Bihlo2012}).
We
search for symmetry generators of the form $Q=\tau(t,x,u)\partial_t+\xi(t,x,u)\partial_x+\eta(t,x,u)\partial_u$
and require that
\begin{equation}\label{c2}
Q^{(5)}\{u_t+u^nu_x+\beta(t)u_{xxx}+\sigma(t)u_{xxxxx}\}=0
\end{equation}
identically, modulo equation~\eqref{eq_gKawahara}. Here  $Q^{(5)}$ is the fifth prolongation of the operator~$Q$~\cite{Olver1986,Ovsiannikov1982}, i.e., in our case
 $Q^{(5)}=Q+\eta^t\partial_{u_t}+\eta^x\partial_{u_x}+\eta^{xxx}\partial_{u_{xxx}}+\eta^{xxxxx}\partial_{u_{xxxxx}}$, where
\begin{gather*}
\eta^x=D_x(\eta)-u_tD_x(\tau)-u_xD_x(\xi),\quad
\eta^{xxx}=D_x(\eta^{xx})-u_{txx}D_x(\tau)-u_{xxx}D_x(\xi),\\
\eta^{xx}=D_x(\eta^x)-u_{tx}D_x(\tau)-u_{xx}D_x(\xi),\quad
\eta^{xxxx}=D_x(\eta^{xxx})-u_{txxx}D_x(\tau)-u_{xxxx}D_x(\xi),\\
\eta^{xxxxx}=D_x(\eta^{xxxx})-u_{txxxx}D_x(\tau)-u_{xxxxx}D_x(\xi),\quad\eta^t=D_t(\eta)-u_tD_t(\tau)-u_xD_t(\xi),
\end{gather*}
$D_t=\partial_t+u_t\partial_{u}+u_{tt}\partial_{u_t}+u_{tx}\partial_{u_x}+\dots{}$ and
$D_x=\partial_x+u_x\partial_{u}+u_{tx}\partial_{u_t}+u_{xx}\partial_{u_x}+\dots{}$
are operators of the total differentiation with respect to~$t$ and~$x$, respectively.
Note that the restriction on $n$ to be integer is inessential for the group classification problem. Therefore, in the course of the study of Lie symmetries $n$ can be assumed as real nonzero constant.

The infinitesimal invariance criterion implies
\[
\tau=\tau(t),\quad
\xi=\xi(t,x), \quad
\eta=\eta^1(t,x)u+\eta^0(t,x),
\]
where $\tau$, $\xi$, $\eta^1$ and $\eta^0$ are arbitrary smooth functions of their variables.
The rest of the determining equations have the form
\begin{gather*}
\eta^1_x=2\xi_{xx},\quad 3(\eta^1_x-\xi_{xx})\beta+5(2\eta^1_{xxx}-\xi_{xxxx})\sigma=0,\\
\tau \sigma_t =(5\xi_x-\tau_t)\sigma,\quad \tau \beta_t =(3\xi_x-\tau_t)\beta +10(\xi_{xxx}-\eta^1_{xx})\sigma,\\
\eta^1_xu^{n+1}+\eta^0_xu^{n}+(\eta^1_t+\eta^1_{xxx}\beta+\eta^1_{xxxxx}\sigma)u+\eta^0_t+\eta^0_{xxx}\beta+\eta^0_{xxxxx}\sigma=0,\\
(\tau_t-\xi_x+n\eta^1)u^n+n\eta^0u^{n-1}+(3\eta^1_{xx}-\xi_{xxx})\beta+(5\eta^1_{xxxx}-\xi_{xxxxx})\sigma-\xi_t=0.
\end{gather*}
The determining equations were verified using the GeM software package~\cite{Cheviakov}.
As $\tau$ and $\sigma$ are functions of $t$ only, the equation $\tau \sigma_t =(5\xi_x-\tau_t)\sigma$ implies
$\xi_{xx}=0$. Then the first determining equation gives $\eta^1_x=0.$
The latter two equations can be split with respect to different powers of $u$. Special cases of the splitting arise if $n=0,1$.
If $n=0$ equations~\eqref{eq_gKawahara} are linear ones and, therefore, excluded from consideration.  The cases  $n\neq1$ and $n=1$ will be investigated separately.

{\rm\bf I.} If $n\neq1$ then the splitting results in $\eta^0=\eta^1_t=\xi_t=0$ and $\tau_t-\xi_x+n\eta^1=0.$
We solve this system together with the derived earlier conditions $\xi_{xx}=\eta^1_x=0$ and get the solution
$\tau=c_1t+c_2,$ $\xi=(c_1+nc_0)x+c_3,$ $\eta^1=c_0,$ $\eta^0=0$,
where $c_i$, $i=0,\dots,3$, are arbitrary constants. Thus, the general form of the infinitesimal generator is
\[Q=(c_1t+c_2)\partial_t+((c_1+nc_0)x+c_3)\partial_x+c_0u\partial_u.\]
The classifying equations on $\beta$ and $\sigma$ are
\begin{equation}\label{c11}
(c_1t+c_2)\beta_t=(2c_1+3nc_0)\beta, \quad (c_1t+c_2)\sigma_t=(4c_1+5nc_0)\sigma.
\end{equation}
To derive
the kernel~$A^{\rm ker}$ of maximal Lie invariance algebras~$A^{\rm max}$ of equations from class~\eqref{eq_gKawahara} (i.e., the Lie invariance algebra admitted by~\eqref{eq_gKawahara} for arbitrary $\beta$ and $\sigma$) we split~\eqref{c11} with respect to $\beta$, $\sigma$ and their derivatives.
Then $c_0=c_1=c_2=0$ and, therefore,  $Q=c_3\partial_x$. Thus, the kernel algebra is the one-dimensional algebra $\langle\partial_x\rangle$.
To get possible extensions of $A^{\rm ker}$ we consider~\eqref{c11} not as two identities but as a system of equations  on $\beta$ and $\sigma$ of the form
 \begin{equation}\label{c12}
(pt+q)\beta_t=r\beta, \quad (pt+q)\sigma_t=\tfrac13(5r+2p)\sigma,
\end{equation}
where $p,$ $q$, and $r$ are arbitrary constants, $p^2+q^2\neq0$.
The equivalence transformations~\eqref{tr_equiv_cor2} act on the coefficients $p,$ $q$, and $r$ of system~\eqref{c12} as follows
\[
\tilde p= \kappa p,\quad \tilde q=\kappa(q\delta_1\delta_3{}^{-n}-p\delta_0),\quad\tilde r=\kappa r,
\]
where $\kappa$ is a nonzero constant. Therefore, there are three inequiva\-lent triples $(p,q,r)$:
$(1,0,\rho)$, $(0,1,1)$ and $(0,1,0)$. We integrate~\eqref{c12} for these values of $(p,q,r)$.
Up to $\hat G^{\sim}_{\alpha=1}$-equivalence $(\beta,\sigma)$ take the values from the set $\left\{\big(\lambda  t^\rho,\delta  t^{\frac{5\rho+2}3}\big),\,\big(\lambda  e^t,\delta  e^{\frac53t}\big),\,(\lambda,\delta)\right\}$.
 Here $\rho$, $\lambda$, and $\delta$
are arbitrary constants with $\lambda\delta\neq0$, $\delta=\pm1\bmod \hat G^{\sim}_{\alpha=1}$.
The respective forms of the infinitesimal generators are
$Q=c_1t\partial_t+\left(\frac{\rho+1}{3}c_1x+c_3\right)\partial_x+\frac{\rho-2}{3n}c_1\partial_u$, $Q=3nc_0\partial_t+\left(nc_0x+c_3\right)\partial_x+c_0\partial_u$, and $Q=c_2\partial_t+c_3\partial_x$, where $c_0,$ $c_1$, $c_2$ and $c_3$ are arbitrary constants.

We have proven the following assertion.
\begin{theorem}
The kernel of the maximal Lie invariance algebras of equations from class~\eqref{eq_gKawahara} {\rm(}resp.~\eqref{eq_ggKawahara}{\rm)} with $n\neq1$
coincides with the one-dimensional algebra $\langle\partial_x\rangle$.
All possible $\hat G^\sim_{\alpha=1}$-inequivalent {\rm(}resp. $G^\sim$-inequivalent{\rm)} cases of extension of the maximal Lie invariance algebras are exhausted
by the cases 1--3 of Table~\ref{TableLieSymG}.
\end{theorem}
\begin{table}[h!]\small \renewcommand{\arraystretch}{1.65}
\begin{center}
\setcounter{tbn}{-1}
\refstepcounter{table}\label{TableLieSymG}
\textbf{Table~\thetable.}
The group classification of the class~$u_t+\alpha u^nu_x+\beta u_{xxx}+\sigma u_{xxxxx}=0$,\quad $n\alpha\beta\sigma\neq0$.
\\[2ex]
\begin{tabular}{|c|c|c|l|}
\hline
&$\beta(t)$&$\sigma(t)$&\hfil Basis of $A^{\max}$ \\
\hline
\multicolumn{4}{|c|}{${n\ne1}$. This case is classified up to $G^{\sim}$-equivalence.}
\\
\hline
0&$\forall$&$\forall$&$\partial_x$\\
\hline
1&
$\lambda t^\rho$&$\delta t^{\frac{5\rho+2}{3}}$&$\partial_x,\,3nt\partial_t+(\rho+1)n x\partial_x+(\rho-2) u\partial_u$\\
\hline
2&$\lambda e^{t}$&$\delta e^{\frac53 t}$&
$\partial_x,\,3n\partial_t+nx\partial_x+u\partial_u$\\
\hline
3&$\lambda $&$\delta$&
$\partial_x,\,\partial_t$\\
\hline
\multicolumn{4}{|c|}{${n=1}$. This case is classified up to $\hat G^{\sim}_{n=1}$-equivalence.}
\\
\hline
$0'$&$\forall $&$\forall$&
$\partial_x,\,t\partial_x+\partial_u$\\
\hline
$1'$&
$\lambda t^\rho$&$\delta t^{\frac{5\rho+2}{3}}$&$\partial_x,\,t\partial_x+\partial_u,\,3t\partial_t+(\rho+1) x\partial_x+(\rho-2) u\partial_u$\\
\hline
$2'$&$\lambda e^{t}$&$\delta e^{\frac53 t}$&
$\partial_x,\,t\partial_x+\partial_u,\,3\partial_t+x\partial_x+u\partial_u$\\
\hline
$3'$&$\lambda $&$\delta$&
$\partial_x,\,t\partial_x+\partial_u,\,\partial_t$\\
\hline
$4'$&$\lambda(t^2+1)^{\frac12} e^{3\nu\arctan t}$&$\delta(t^2+1)^{\frac32} e^{ 5\nu\arctan t}$&
$\partial_x,\,t\partial_x+\partial_u,\,$\\
&&&$(t^2+1)\p_t+(t+\nu)x\p_x+((\nu-t)u+x)\p_u$\\
\hline
\end{tabular}
\\[2ex]
\parbox{150mm}{Here $\alpha=1\bmod\, G^\sim$,  $\rho$ and $\nu$ are arbitrary constants, $\rho\geqslant1/2$, $\nu\geqslant0$; $\delta$ and $\lambda$ are nonzero constants, $\delta=\pm1\bmod\, G^\sim.$}
\end{center}
\end{table}

{\rm\bf II.} If $n=1$ then the determining equations lead to the system $\eta^1_x=\xi_{xx}=0$, $\eta^0=\xi_t,$ $\tau_t-\xi_x+\eta^1=0,$ $\eta^0_x+\eta^1_t=0$, $\eta^0_t=\eta^0_{xx}=0$, $\tau\beta_t=(3\xi_x-\tau_t)\beta,$ and $\tau\sigma_t=(5\xi_x-\tau_t)\sigma.$
We solve firstly the equations that do not contain arbitrary elements and get the form of the infinitesimal generator
\[
Q=(c_2t^2+2c_1t+c_0)\partial_t\!+\!((c_2t+c_1+c_3)x+c_4t+c_5)\partial_x\!+\!((c_3-c_1 -c_2t)u+c_2x+c_4)\partial_u,
\]
where $c_i$, $i=0,\dots,5$, are arbitrary constants.
The system of classifying equations takes the form
\begin{equation*}
(c_2t^2+2c_1t+c_0)\beta_t=(c_2t+c_1+3c_3)\beta,\quad
(c_2t^2+2c_1t+c_0)\sigma_t=(3c_2t+3c_1+5c_3)\sigma.
\end{equation*}
If $\beta$ and $\sigma$ are arbitrary we can split the latter equations with respect to them and their derivatives. As a result we obtain that
$c_3=c_2=c_1=c_0=0$. Therefore, $Q=(c_4t+c_5)\partial_x+c_4\partial_u$ and
the kernel $A^{\rm ker}$ of the maximal Lie invariance algebras of equations from class~\eqref{eq_gKawahara_n1}
coincides with the two-dimensional algebra $\langle\p_x,\,t\p_x+\p_u\rangle$.

The group classification of class~\eqref{eq_gKawahara_n1} is equivalent to the integration of the classifying equations
up to the $G^{\sim}_{\alpha=n=1}$-equivalence.
Combined with multiplication by a nonzero constant,
each transformation from the equivalence group~$G^{\sim}_{\alpha=n=1}$ can be  extended to the coefficient quadruple $(p,q,r,s)$
of the system
\begin{equation}\label{cl_syst}
(p t^2+q t+r)\beta_t=(p t+s)\beta,\quad(p t^2+q t+r)\sigma_t=(3p t+(5s+2q)/3)\sigma,
\end{equation}
where $p,$ $q$, $r$ and $s$ are arbitrary constants, $p^2+q^2+r^2\neq0$,
in the following way
\begin{gather*}
\begin{array}{l}
\tilde p=\kappa(p d^2-q c d+r c^2),\quad
\tilde q=\kappa(-2p b d+q(a d+b c)-2r a c),
\\[1ex]
\tilde r=\kappa(p b^2-q a b +r a^2),
\quad\tilde s=\kappa(r a c+q b c-p b d+ s\Delta),
\end{array}
\end{gather*}
where $\Delta=a d - b c$ and $\kappa$ is an arbitrary nonzero constant.

It can be proved that
there are only three $G^{\sim}_{\alpha=n=1}$-inequivalent values of the triple $(p,q,r)$
depending upon the sign of $D=q^2-4pr$,
\begin{gather*}
(0,1,0)\quad\mbox{if}\quad D>0, \quad
(0,0,1)\quad\mbox{if}\quad  D=0, \quad \mbox{\rm and} \quad
(1,0,1)\quad\mbox{if}\quad  D<0.
\end{gather*}
The technique of the proof can be found in~\cite{vane2012b}.
The remaining task is to consider whether there is possibility
to scale the constant $s$ in each of the three distinct cases for $(p,q,r)$. As a result we get the following statement.

\begin{proposition}\label{LemmaOntransOfCoeffsOfClassifyingSystem2}
Up to $G^{\sim}_{\alpha=n=1}$-equivalence
the parameter quadruple~$(p,q,r,s)$ can be assumed to belong to the set
\[
\{(0,1,0,\rho),\ (0,0,1,1),\ (0,0,1,0),\ (1,0,1,\bar s)\},
\]
where $\bar s$ is an arbitrary constant, $\rho\geqslant\frac12$, $\bar s\geqslant 0.$
\end{proposition}
\begin{table}[t!]\small \renewcommand{\arraystretch}{1.85}
\begin{center}
\setcounter{tbn}{-1}
\refstepcounter{table}\label{TableLieSymGext}
\textbf{Table~\thetable.}
The group classification of the class~$u_t+\alpha u^nu_x+\beta u_{xxx}+\sigma u_{xxxxx}=0$,\quad $n\alpha\beta\sigma\neq0$, \\using no equivalence.
\\[2ex]
\begin{tabular}{|@{\,}c@{\,}|@{\,}c@{\,}|@{\,}c@{\,}|@{\,\,}l@{\,}|}
\hline
&$\beta(t)$&$\sigma(t)$&\hfil Basis of $A^{\max}$ \\
\hline
\multicolumn{4}{|c|}{${n\ne1}$}
\\
\hline
0&$\forall$&$\forall$&$\partial_x$\\
\hline
1&
$\lambda_1\alpha(T+l)^\rho$&$\lambda_2\alpha(T+l)^{\frac{5\rho+2}{3}}$&$\partial_x,\,3n(T\!+\!l)\alpha^{-1}\partial_{t}\!+\!n(\rho\!+\!1)x\partial_{x}\!+\!(\rho\!-\!2)u\partial_{u}$\\
\hline
2&$\lambda_1\alpha e^{mT}$&$\lambda_2\alpha e^{\frac53 mT}$&
$\partial_x,\, 3n\alpha^{-1}\partial_{t}+nmx\partial_{x}+
  mu\partial_{u}$\\
\hline
3&$\lambda_1\alpha $&$\lambda_2\alpha$&
$\partial_x,\,\alpha^{-1}\partial_{t}$\\
\hline
\multicolumn{4}{|c|}{${n=1}$}
\\
\hline
$0'$&$\forall $&$\forall$&
$\partial_x,\,T\partial_x+\partial_u$\\
\hline
$1'$&
$\lambda_1\alpha(aT\!+\!b)^\rho(cT\!+\!d)^{1-\rho}$&$\lambda_2\alpha(aT\!+\!b)^{\frac{5\rho+2}{3}}(cT\!+\!d)^{\frac{7-5\rho}{3}}$&
$\partial_x,\,T\partial_x+\partial_u, 3(aT\!+\!b)(cT\!+\!d)\alpha^{-1}\partial_{t}+
  $\\
  &&&$\left(3acT\!+\!ad(\rho\!+\!1)+bc(2\!-\!\rho)\!\right)x\partial_{x}+$\\
   &&&$\left(3acx\!-\!(3acT\!+\!ad(2\!-\!\rho)\!+\!bc(\rho\!+\!1)) u\!\right)\partial_{u}$\\
\hline
$2'$&$\lambda_1\alpha(cT\!+\!d)\exp\!\left(\frac{aT+b}{cT+d}\right)$&$\lambda_2\alpha(cT\!+\!d)^3\exp\!\left(\frac53 \frac{aT+b}{cT+d}\right)$&
$\partial_x,\,T\partial_x+\partial_u,\,3(cT\!+\!d)^2\alpha^{-1}\partial_{t}+$\\
 &&&$\left(3c(cT\!+\!d)\!+\!\Delta\right)x\partial_{x}+$\\
  &&&$
 \left(3c^2x\!+\! (\Delta\!-\!3c(cT\!+\!d))u\right)\partial_{u}$\\
\hline
$3'$&$\lambda_1\alpha(cT\!+\!d)$&$\lambda_2\alpha(cT\!+\!d)^3$&
$\partial_x,\,T\partial_x+\partial_u,\,(cT\!+\!d)^2\alpha^{-1}\partial_{t}+$\\
  &&&$c(cT\!+\!d)x\partial_{x}+
  c(c x\!-\!(cT\!+\!d) u)\partial_{ u}$\\
\hline
&$\lambda_1\alpha\exp\!\left(3\nu\arctan\frac{aT+b}{cT+d}\right)\!\!\times$&$\lambda_2\alpha\exp\!\left(5\nu\arctan\frac{aT+b}{cT+d}\right)\!\!\times$&
$\partial_x,\,T\partial_x+\partial_u,\,\left(\!(aT\!+\!b)^2\!+\!(cT\!+\!d)^2\!\right)\!\alpha^{-1}\partial_{t}+
$\\
$4'$&$
\left((aT\!+\!b)^2\!+\!(cT\!+\!d)^2\right)^{\frac12}$&$\left(\!(aT\!+\!b)^2\!+\!(cT\!+\!d)^2\right)^{\frac32}$&
$  \left(a(aT\!+\!b)\!+\!c(cT\!+\!d)\!+\!\Delta\nu\right)x\partial_{x}+
  $\\
  &&&$\left(\!(a^2\!+\!c^2) x\!-\!(a(aT\!+\!b)\!+\!c(cT\!+\!d)\!-\!\Delta\nu) u\!\right)\!\partial_{u}$\\
\hline
\end{tabular}
\\[1.5ex]
\parbox{155mm}{Here  $\lambda_1$, $\lambda_2$, $a$, $b$, $c$, $d$, $l$, $m$, $\rho$ and $\nu$ are arbitrary constants, $\lambda_1\lambda_2(c^2+d^2)\ne0$, $\Delta=ad-bc\neq0$, $\alpha$ is an arbitrary nonvanishing smooth function of $t$, $T=\int\! \alpha(t) {\rm d}t$.}
\end{center}
\end{table}

We integrate~\eqref{cl_syst} for the values of $(p,q,r,s)$ presented in Proposition~1, then substitute the derived forms of $\beta$ and $\sigma$ to the classifying
equations in order to get $c_i$, $i=0,\dots,3,$ (the constants $c_4$ and $c_5$ are arbitrary). We get that
all $G^\sim_{\alpha=n=1}$-inequivalent cases of Lie symmetry extension are exhausted by the following:

\medskip
\noindent
$(\beta,\sigma)=\big(\lambda  t^\rho,\delta  t^{\frac{5\rho+2}3}\big),\ \rho\geqslant\frac12\colon$
$Q=2c_1t\p_t+\left(\frac{2}{3}(\rho+1)c_1x+c_4t+c_5\right)\p_x+\left(\frac{2}{3}(\rho-2)c_1u+c_4\right)\p_u$;

\medskip

\noindent
$(\beta,\sigma)=\big(\lambda  e^t,\delta  e^{\frac53t}\big)\colon$
$Q=3c_3\p_t+(c_3x+c_4t+c_5)\p_x+(c_3u+c_4)\p_u$;

\medskip
\noindent
$(\beta,\sigma)=\left(\lambda,\delta\right)$:
$Q=c_0\partial_t+(c_4t+c_5)\partial_x+c_4\partial_u$;

\medskip
\noindent
$(\beta,\sigma)=\big(\lambda(t^2+1)^{\frac12} e^{3\nu\arctan t},\delta(t^2+1)^{\frac32} e^{5 \nu\arctan t}\big)$, $\nu\geqslant0$ ($\nu:=\bar s/3$):

\smallskip

$Q=c_2(t^2+1)\partial_t+(c_2(t+\nu)x+c_4t+c_5)\partial_x+\left(c_2((\nu-t)u+x)+c_4\right)\partial_u$.

\medskip

In all four adduced cases the maximal Lie invariance algebras are three-dimensional.
The following assertion is true.
\begin{theorem}
The kernel of the maximal Lie invariance algebras of equations from class~\eqref{eq_gKawahara_n1} {\rm(}resp.~\eqref{eq_Kawahara_n=1}{\rm)}
coincides with the two-dimensional algebra $\langle\partial_x,\,t\p_x+\p_u\rangle$.
All possible $G^\sim_{\alpha=n=1}$-inequivalent {\rm(}resp. $\hat G^\sim_{n=1}$-inequivalent{\rm)} cases of extension of the maximal Lie invariance algebras are exhausted
by the cases $1'$--$\,4'$ of Table~\ref{TableLieSymG}.
\end{theorem}

\vspace{-1ex}

To derive the complete list of Lie symmetry extensions for the entire class~\eqref{eq_ggKawahara},
where arbitrary elements are not simplified by point transformations, we use the equivalence-based approach~\cite{Vaneeva2012}.
The results are collected in Table~2.

The presented group classification reveals equations of the form~\eqref{eq_ggKawahara} that may be of interest for applications and  for which the classical  Lie reduction method can be  used.

\section{Symmetry reductions and exact solutions}

The Lie symmetry operators derived as a result of solving the group classification problem can
be applied to construction of exact solutions of the corresponding equations. The reduction method
with respect to subalgebras of Lie invariance algebras is algorithmic  and
well-known; we refer to the classical textbooks on the subject~\cite{Olver1986,Ovsiannikov1982}.
In order to get an optimal system of group-invariant solutions reductions should be performed with respect to subalgebras from the optimal system~\cite[Section~3.3]{Olver1986}.

Consider firstly the structure of the two and three-dimensional Lie algebras spanned by the generators presented in Table 1, using notations of~\cite{pate1977a}.
In Cases 1--3 and Case $0'$ the maximal Lie-invariance algebras are two-dimensional. In Case $0'$, Case 1 with $\rho=-1$, and Case 3 they are Abelian ($2A_1$). The algebras adduced in Case 1 with $\rho\neq-1$ and Case 2 are non-Abelian ($A_2$). The algebras with basis operators presented in Cases $1'$--$4'$ are three-dimensional. In Case~$1'$ with $\rho\ne-1,2$ the maximal Lie invariance algebra is of the type $A_{3.4}$ if $\rho=1/2$, $A^a_{3.5}$ with  $a=\frac{\rho-2}{\rho+1}$ or $a=\frac{\rho+1}{\rho-2}$ if $\rho>1/2$ or $\rho<1/2$, respectively.
If $\rho=-1$ or $\rho=2$, then $A^{\rm max}$ from Case~$1'$ is $A_1\oplus A_2$. In other cases the maximal Lie invariance algebras are of the following types:
Case~$2'$ --- $A_{3.2}$,
Case~$3'$ --- the Weyl algebra $A_{3.1}$, Case~$4'$ --- $A^a_{3.7}$ with $a=|\nu|$.

\begin{table}[b!]\small \renewcommand{\arraystretch}{1.65}
\begin{center}
\textbf{Table 3.} Optimal systems of one-dimensional subalgebras of $A^{\rm max}$  presented in Table 1.
\\[2ex]
\begin{tabular}
{|c|l|}
\hline
\hfil Case
&
\hfil Optimal system
\\
\hline
$1_{\rho\neq-1}$
&
${\mathfrak g}^{\,}_0=\langle\partial_x\rangle,
\quad
{\mathfrak g}^{\,}_{1.1}=\langle 3nt\partial_t+(\rho+1)n x\partial_x+(\rho-2) u\partial_u\rangle$
\\
\hline
$1_{\rho=-1}$
&
${\mathfrak g}^{\,}_0=\langle\partial_x\rangle,
\quad
{\mathfrak g}^a_{1.2}=\langle nt\partial_t+a\partial_x-u\partial_u\rangle$
\\
\hline
2
&
${\mathfrak g}^{\,}_0=\langle\partial_x\rangle,
\quad
{\mathfrak g}^{\,}_2=\langle3n\partial_t+nx\partial_x+u\partial_u\rangle$
\\
\hline
3
&
${\mathfrak g}^{\,}_0=\langle\partial_x\rangle,
\quad
{\mathfrak g}^a_{3}=\langle\partial_t+a\partial_x\rangle$
\\
\hline
$0'$
&
${\mathfrak g}^{\,}_0=\langle\partial_x\rangle,
\quad
{\mathfrak g}^a_{0'}=\langle (t+a)\partial_x+\partial_u\rangle$
\\
\hline
${1'}_{\rho\neq-1,2}$
&
${\mathfrak g}^{\,}_0=\langle\partial_x\rangle,
\quad
{\mathfrak g}^{\sigma}_{0'}=\langle (t+\sigma)\partial_x+\partial_u\rangle\quad {\mathfrak g}_{1'.1}=\langle 3t\partial_t+(\rho+1) x\partial_x+(\rho-2) u\partial_u\rangle$
\\
\hline
${1'}_{\rho=-1}$
&
$\mathfrak g^{\,}_0=\langle\partial_x\rangle,
\quad
\mathfrak g^{\sigma}_{0'}=\langle(t+\sigma)\partial_x+\partial_u\rangle,
\quad
\mathfrak g^a_{1'.2}=\langle t\partial_t+a\partial_x-u\partial_u\rangle$, \\
\hline

${1'}_{\rho=2}$
&
${\mathfrak g}^{\,}_0=\langle\partial_x\rangle,
\quad
{\mathfrak g}^{\sigma}_{0'}=\langle(t+\sigma)\partial_x+\partial_u\rangle,
\quad
{\mathfrak g}^a_{1'.3}=\langle t\partial_t+\left(x+at\right)\partial_x+a\partial_u\rangle,$
\\
\hline
$2'$
&
$\mathfrak g^{\,}_0=\langle\partial_x\rangle,
\quad
\mathfrak g_{0'}=\langle t\partial_x+\partial_u\rangle,
\quad
\mathfrak g^{\,}_{2'}=\langle3\partial_t+x\partial_x+u\partial_u\rangle$, \\
\hline
$3'$
&
$\mathfrak g^{\,}_0=\langle\partial_x\rangle,
\quad
\mathfrak g^{\,}_{3'.1}=\langle\partial_t\rangle,\quad\mathfrak g^a_{3'.2}=\langle a\partial_t+2t\partial_x+2\partial_u\rangle$
\\
\hline
$4'$
&
$\mathfrak g^{\,}_0=\langle\partial_x\rangle,
\quad
\mathfrak g^{\,}_{4'}=\langle(t^2+1)\partial_t+(t+\nu)x\partial_x+(x+(\nu-t)u)\partial_u\rangle$
\\
\hline
\end{tabular}
\end{center}
In all cases $a\in\mathbb R$, $n\ne0$,  $\sigma\in\{-1,0,1\}$.
\end{table}

If a one-dimensional invariance algebra is spanned by an operator  $Q=\tau\partial_t+\xi\partial_x+\eta\partial_u$,
then the associated ansatz reducing the corresponding PDE with two independent variables to an ODE is found as a solution of the invariant surface condition $Q[u]:=\tau u_t+\xi u_x-\eta=0$.
In practice the related characteristic system $\frac{{\rm d}t}{\tau}=\frac{{\rm d}x}{\xi}=\frac{{\rm d}u}{\eta}$
has to be solved.
Ansatzes and reduced equations obtained for equations from class~\eqref{eq_gKawahara}
using one-dimensional subalgebras from Table~3 are collected in Table~4.
Reductions associated with the subalgebra $\mathfrak g_0$ are not considered
since they lead to constant solutions only.
We do not present  reductions with respect to the subalgebras $\mathfrak g_{1'.1}$, $\mathfrak g^a_{1'.2}$ and $\mathfrak g_{2'}$ since these subalgebras are specifications of the subalgebras
 $\mathfrak g_{1.1}$, $\mathfrak g^a_{1.2}$ and $\mathfrak g_2$ for the case $n=1$.
The reduction for the case $1'_{\rho=2}$ is not performed because this case is equivalent to $1'_{\rho=-1}$. Indeed,  the equations $u_t+uu_x+\lambda t^2u_{xxx}+\delta t^4 u_{xxxxx}=0$ and $u'_{t'}+u'{u'}_{x'}+\lambda/ {t'}{u'}_{x'x'x'}+\delta/{t'} {u'}_{x'x'x'x'x'}=0$
are linked by the transformation $t'=1/t$, $x'=-x/t$, $u'=tu-x$.

\begin{table}[b!]\small \renewcommand{\arraystretch}{2}
\begin{center}
\textbf{Table 4.} Similarity reductions of the equations~$u_t+u^nu_x+\beta(t)u_{xxx}+\sigma(t)u_{xxxxx}=0$.
\\[2ex]
\begin{tabular}
{|c|c|c|c|l|}
\hline
Case&$\mathfrak g$& $\omega$ &\hfil Ansatz, $u=$ &\hfil Reduced ODE
\\
\hline
\multicolumn{5}{|c|}{Reductions for arbitrary nonzero $n$}
\\
\hline
$1_{\rho\ne-1}$&${\mathfrak g}^{\,}_{1.1}$&$xt^{-\frac{\rho+1}{3}}$ &
$t^{\frac{\rho-2}{3n}}\varphi(\omega)$ & $\delta\varphi'''''+ \lambda\varphi'''+\left(\varphi^n-\frac{\rho+1}{3}\omega\right)\varphi'+\frac{\rho-2}{3n}\varphi=0$
\\
\hline
$1_{\rho=-1}$ & ${\mathfrak g}^a_{1.2}$&$x-\frac{a}{n}\ln t$ & $t^{-\frac{1}{n}}\varphi(\omega)$ & $\delta\varphi'''''+ \lambda
\varphi'''+\left(\varphi^n-\frac{a}{n}\right)\varphi'-\frac1n\varphi=0$
\\
\hline
2 &${\mathfrak g}_{2}$& $xe^{-\frac13t}$ & $e^{\frac1{3n}t}\varphi(\omega)$ & $
\delta\varphi'''''+\lambda\varphi'''+\left(\varphi^n-\frac1{3}{\omega}\right)\varphi'+\frac1{3n}\varphi=0$
\\
\hline
3 &${\mathfrak g}^a_{3}$ &$x-at$ & $\varphi(\omega)$ & $
\delta\varphi'''''+\lambda\varphi'''+\left(\varphi^n-a\right)\varphi'=0$
\\
\hline
\multicolumn{5}{|c|}{Specific reductions for $n=1$}
\\
\hline
$0'$ &${\mathfrak g}^a_{0'}$& $t$ & $\varphi(\omega)+\dfrac{x}{t+a}$ & $(\omega+a)\varphi'+\varphi=0$
\\
\hline
$3'$ &${\mathfrak g}^a_{3'.1}$& $x$ & $\varphi(\omega)$ & $\delta\varphi'''''+\lambda\varphi'''+\varphi\varphi'=0$
\\
\hline
$3'$ &${\mathfrak g}^a_{3'.2}$& $x-{t^2}/a$ & $2t/a+\varphi(\omega),\,\,\,a\neq0$ & $\delta\varphi'''''+\lambda\varphi'''+\varphi\varphi'+2/a=0$
\\
\hline
$4'$ &${\mathfrak g}^{\,}_{4'}$& $\dfrac{xe^{-\nu\arctan t}}{\sqrt{t^2+1}}$ &
$\dfrac{e^{\nu\arctan t}}{\sqrt{t^2+1}}\varphi(\omega)+\dfrac{xt}{t^2+1}$ & $\delta\varphi'''''+\lambda \varphi'''+(\varphi-\nu\omega)\varphi'+\nu
\varphi+\omega=0$
\\[1.5mm]
\hline
\end{tabular}
\end{center}
Here $a$ is an arbitrary constant.
\end{table}
\vspace{-1ex}

The first-order reduced equation from Table~4,
$(\omega+a)\varphi'+\varphi=0$,
gives the ``degenerate'' solution of~\eqref{eq_gKawahara} for arbitrary values of $\beta(t)$ and $\sigma(t)$,
$
u=(x+c)/(t+a),
$
where $c$ and $a$ are arbitrary constants.
Using transformation~\eqref{eq_gauge} we get the ``degenerate'' exact solution of equation~\eqref{eq_Kawahara_n=1} in the form
\begin{gather}
\label{solution5}
u=\frac{x+c}{\int\! \alpha(t) {\rm d}t+a}.
\end{gather}

Consider fifth-order reduced ODEs from Table~4. Cases 3 and $3'$ correspond to the constant-coefficient generalized Kawahara equations.
The corresponding ODEs were heavily studied in the literature, see, e.g.,~\cite{Bagderina2008,Demina&Kudryashov2010,Kudryashov2012,Parkes&Duffy} and references therein. We concentrate our attention on variable coefficient cases.

\subsection{Exact solutions for equations reducible to their constant coefficients counterparts}
In recent papers~\cite{Kaur&Gupta,Wazwaz2010} different techniques for finding exact solutions were applied to construct exact solutions of
Kawahara equations with time-dependent coefficients. In both papers exact solutions were derived for equations whose coefficients obey
additional constraints, namely, when all the coefficients are proportional to each other. Theorem~3 implies that such variable coefficient equations from class~\eqref{eq_ggKawahara} are reducible to  constant coefficient Kawahara equations.

In our opinion the optimal way to get exact solutions for equations from~\eqref{eq_ggKawahara} that are reducible to the constant-coefficient equations from this class is to take known solutions
for constant coefficient equations and then to make  a corresponding change of variables. In such a way it is possible to construct exact solution not only
for the case when  the coefficients in~\eqref{eq_ggKawahara}  are proportional but also (if $n=1$) for equations of the form~\eqref{eq_ggKawahara} whose coefficients satisfy conditions~\eqref{criterion2}.

We derive the corresponding changes of variables using Theorem~1 for the case $n\neq1$ and Theorem~2 for the case $n=1$. The following statement is true. The equations from class~\eqref{eq_ggKawahara}
\begin{gather}\label{eq_rKawahara1}
\textstyle u_t+\alpha(t)u^nu_x+\tilde \beta\alpha(t)u_{xxx}+\tilde \sigma\alpha(t)u_{xxxxx}=0,\quad\mbox{and},\\\label{eq_rKawahara2}
\textstyle u_t+\alpha(t)uu_x+\tilde \beta\alpha(t)(\delta_3\int\!\alpha(t){\rm d}t+\delta_4)u_{xxx}+\tilde \sigma\alpha(t)(\delta_3\int\!\alpha(t){\rm d}t+\delta_4)^3u_{xxxxx}=0,
\end{gather}
where $\alpha(t)$ is a smooth nonvanishing function,
reduce to the constant coefficient Kawahara equations
\begin{gather}
\label{eq_ggKawahara01}
\tilde u_{\tilde t}+\tilde \alpha \tilde u^n {\tilde u}_{\tilde x} +\tilde \beta \tilde u_{\tilde x\tilde x\tilde x}+\tilde \sigma {\tilde u}_{\tilde x\tilde x\tilde x\tilde x\tilde x}=0,\quad\mbox{and}\\
\label{eq_ggKawahara0}
\tilde u_{\tilde t}+\tilde \alpha \tilde u {\tilde u}_{\tilde x} +\tilde \beta \tilde u_{\tilde x\tilde x\tilde x}+\tilde \sigma {\tilde u}_{\tilde x\tilde x\tilde x\tilde x\tilde x}=0
\end{gather}
via the transformations
\begin{gather}\nonumber
\textstyle \tilde t=\int\!\alpha(t){\rm d}t,\quad\tilde x= x,\quad\tilde u=\tilde\alpha^{-\frac1n} u,\quad\mbox{and}\\[2ex]\arraycolsep=0ex\label{tr}
 \begin{array}{l}
 \tilde t=\int\!\! {\alpha(t)}{(\delta_3\int\!\alpha(t){\rm d}t+\delta_4)^{-2}}{\rm d}t,\quad\tilde x={(x+\delta_1)}{(\delta_3\int\!\alpha(t){\rm d}t+\delta_4)^{-1}},\\[2ex]\tilde u=\textstyle{\left(\!(\delta_3\int\!\alpha(t){\rm d}t+\delta_4)u-(x+\delta_1)\delta_3\!\right)}/{\tilde\alpha},
\end{array}
\end{gather}
respectively.
Here $\delta_i$, $i=1,3,4$, $\tilde\alpha$, $\tilde\beta$, and $\tilde\sigma$ are arbitrary constants with $\tilde\alpha\tilde\beta\tilde\sigma(\delta_3^2+\delta_4^2)\neq0.$

We take a family of solitary wave solutions of the Kawahara equation~\eqref{eq_ggKawahara0} of the form
\begin{gather*}
\tilde u =-\frac{264992 \tilde \sigma^2 \kappa^5 - 7280 \tilde\beta\tilde \sigma \kappa^3-31 \tilde \beta^2 k+507\tilde \sigma\mu}{ 507 \tilde \alpha\tilde \sigma\kappa}\\
\phantom{u=\,}-\frac{280\kappa^2(\tilde \beta-104\tilde \sigma\kappa^2)}{13\tilde \alpha}\tanh^2(\kappa \tilde x+\mu\tilde t+\chi)-\frac{1680\tilde \sigma\kappa^4}{\tilde \alpha}\tanh^4(\kappa \tilde x+\mu\tilde t+\chi)
\end{gather*}
with $\kappa$  given by
\[\kappa_{1,2}=\pm\tfrac{\sqrt{-13\tilde \beta\tilde\sigma}}{26\tilde\sigma},\quad\kappa_{3,4}=\pm\tfrac{\sqrt{65\tilde\beta\tilde\sigma(31-3 i\sqrt{31})}}{260\tilde\sigma},\quad
\kappa_{5,6}=\pm\tfrac{\sqrt{65\tilde\beta\tilde\sigma(31+3 i\sqrt{31})}}{260\tilde\sigma},\]
  $\mu$ and $\chi$ being arbitrary constants~\cite{Kudryashov2012}. The corresponding exact solution of~\eqref{eq_rKawahara2}, derived with the usage of~\eqref{tr}, is
\begin{gather*}
u=\frac1{\delta_3\int\!\alpha(t){\rm d}t+\delta_4}\bigg(\delta_3(x+\delta_1)-\frac{264992\, \tilde \sigma^2 \kappa^5 - 7280 \tilde\beta\tilde \sigma \kappa^3-31 \tilde \beta^2 \kappa+507\tilde \sigma\mu}{ 507 \tilde \sigma\kappa}\\
\phantom{u=\,}-\frac{280}{13}\kappa^2(\tilde \beta-104\tilde \sigma\kappa^2)\tanh^2(\kappa \tilde x+\mu\tilde t+\chi)-{1680\tilde \sigma\kappa^4}\tanh^4(\kappa \tilde x+\mu\tilde t+\chi)\bigg),
\end{gather*}
where $\tilde t=\int\!\! {\alpha(t)}{(\delta_3\int\!\alpha(t){\rm d}t+\delta_4)^{-2}}{\rm d}t,$ $\tilde x={(x+\delta_1)}{(\delta_3\int\!\alpha(t){\rm d}t+\delta_4)^{-1}},$  $\delta_1$,  $\mu$ and $\chi$ are arbitrary constants, $\kappa$ takes the six values adduced above.

A family of solutions for equation~\eqref{eq_rKawahara1} with $n=2$ has the form
\begin{equation}\label{sol2}
u=\frac{40k^2\tilde\sigma-\tilde\beta}{\sqrt{-10\tilde \sigma}}+6k^2\sqrt{-10\tilde \sigma}\tanh^2\left(k x+\frac{k}{10\tilde \sigma}(240k^4\tilde\sigma^2+\tilde \beta^2)\textstyle{\int}\!\alpha(t){\rm d}t+\chi\right),
\end{equation}
where $k$ and $\chi$ are arbitrary constants with $k\neq0$.  On Figs.~1--3 we present the graphs of solution~\eqref{sol2} for certain values of parameters and different time inhomogeneities.

\begin{figure}[h!]
\begin{minipage}[t]{50mm}
\centering
\includegraphics[width=45mm]{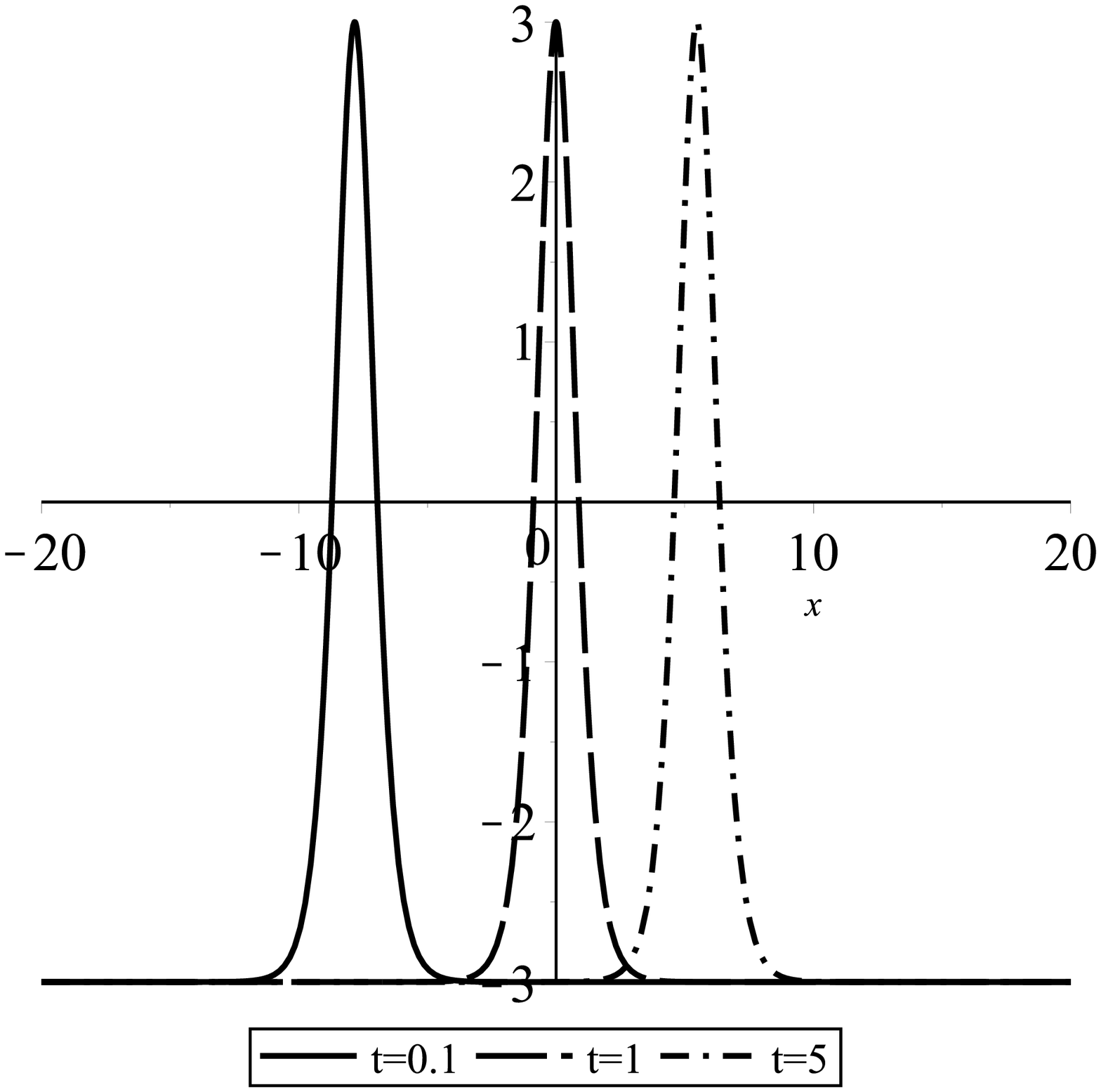}
\caption{\small Solution~\eqref{sol2} for $\alpha(t) = 1/t$, $\sigma = -0.1$, $\beta = -1$, $k = 1$, $\chi = 0$.}
\end{minipage}
\quad
\begin{minipage}[t]{50mm}
\centering
\includegraphics[width=45mm]{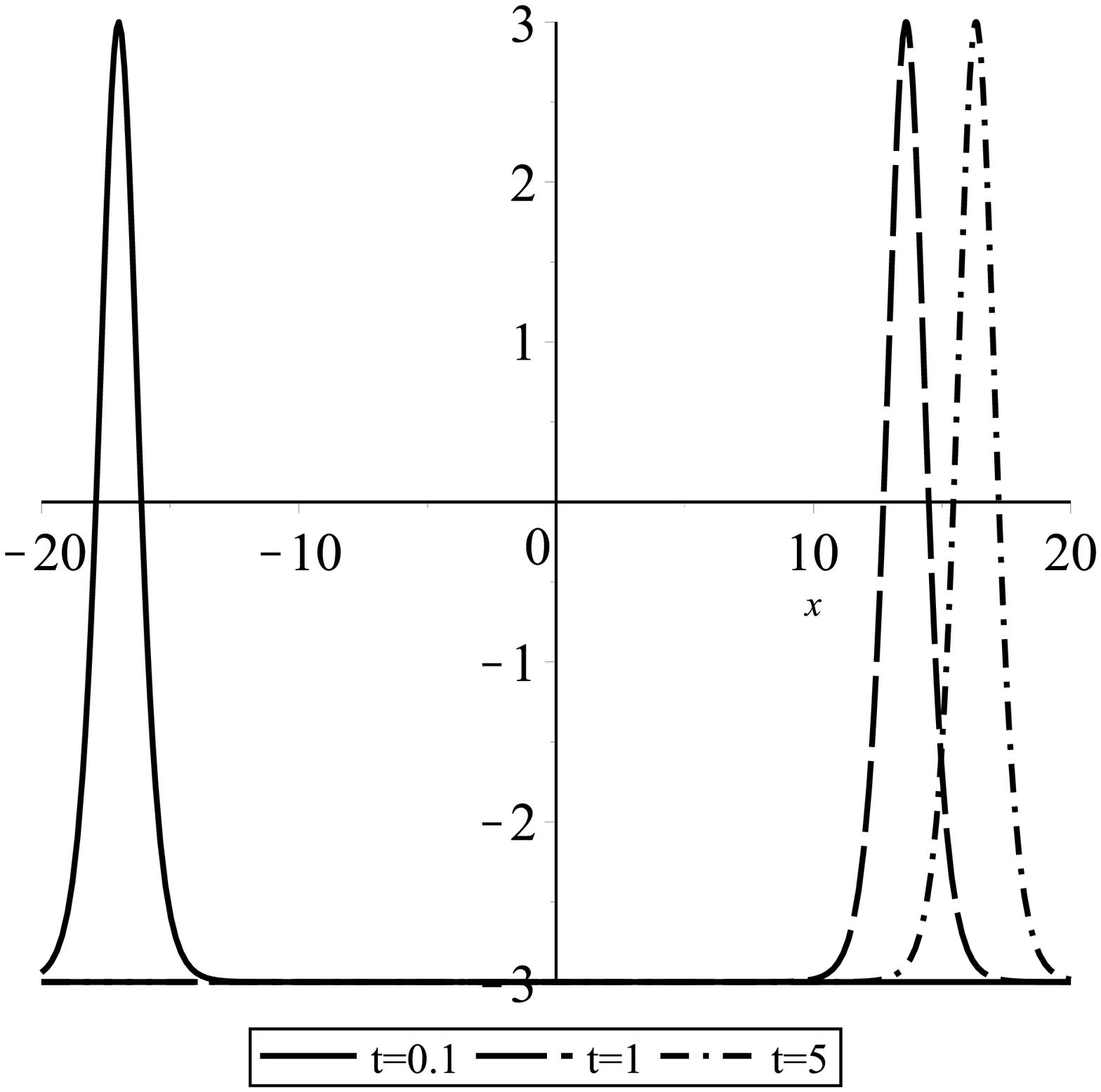}
\caption{\small Solution~\eqref{sol2} for $\alpha(t) = 1/t^2$, $\sigma = -0.1$,
$\beta = -1$, $k = 1$, $\chi = -17$.
}
\end{minipage}
\quad
\begin{minipage}[t]{50mm}
\centering
\includegraphics[width=45mm]{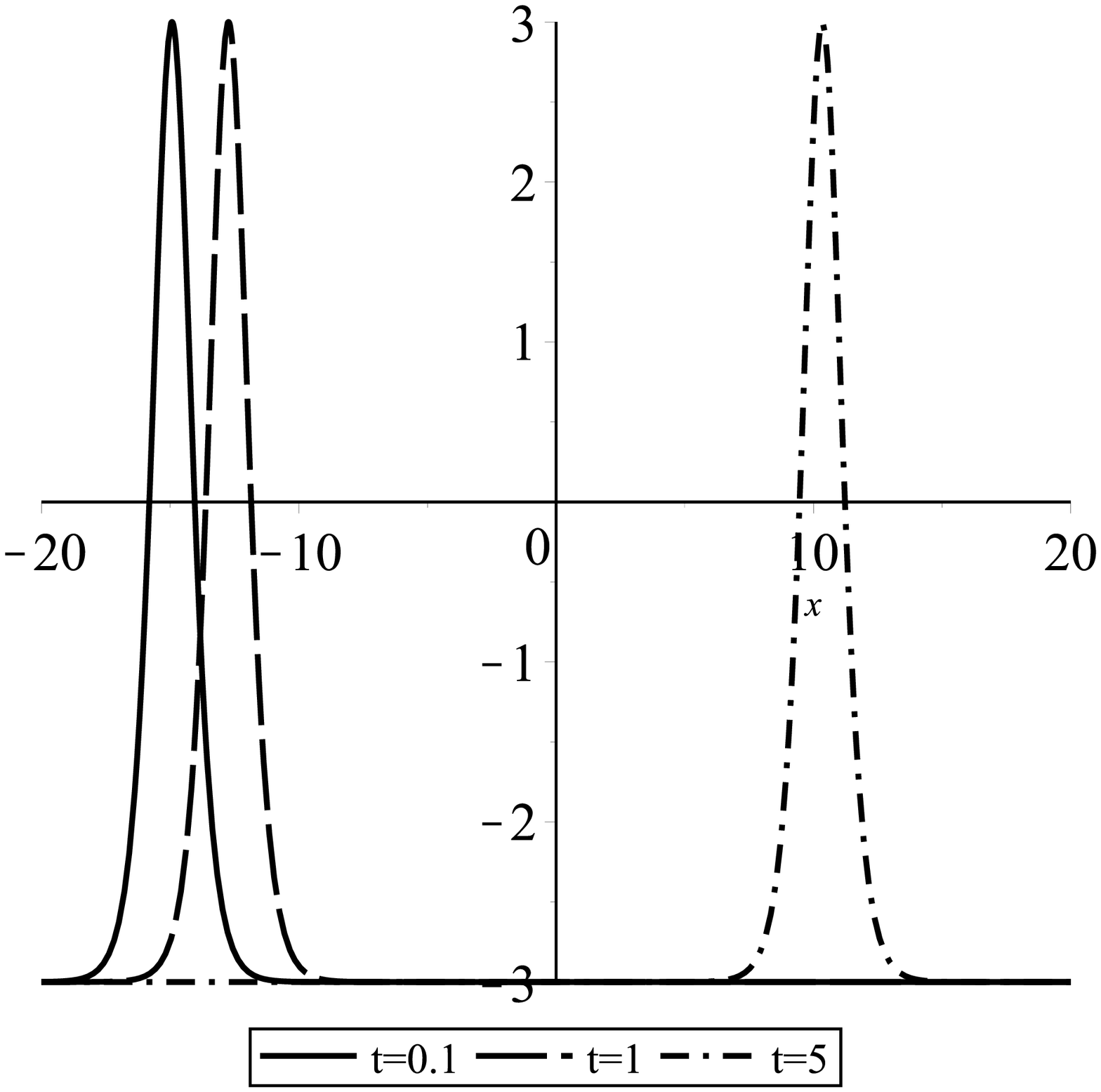}
\caption{\small Solution~\eqref{sol2} for $\alpha(t) = \sqrt{t}$, $\sigma = -0.1$,
$\beta = -1$, $k = 1$, $\chi = 15$.
}
\end{minipage}
\end{figure}

\subsection{Numerical solutions using Lie symmetries}

Exact solutions of the fifth-order ODEs presented in  Cases $1$, $2$ and $4'$ of Table~4 are not known.
At the same time behavior of solutions for variable coefficients models is what we are most interested in.
The Lie reductions obtained can be useful in seeking solutions of equations~\eqref{eq_ggKawahara}
accompanied with boundary conditions that are invariant with respect to the corresponding Lie symmetry algebras~\cite{Bluman&Anco2002}.

Consider a class of boundary value problems (BVPs) for variable coefficient generalized Kawahara equations,
\begin{gather}\label{BV_Kawahara}
u_t+u^nu_x+\lambda t^{\rho}u_{xxx}+\delta t^{\frac{5\rho+2}3}u_{xxxxx}=0,\quad t>t_0,\quad x>0,\quad n\in\mathbb{N}, 
\\[1ex]\label{BV_Kawahara_bc}
u(t,0)=\gamma_0 t^{\frac{\rho-2}{3n}},\quad
\dfrac{\partial^{\,i}u(t,x)}{\partial x^i}\bigg|_{x=0}=\gamma_i t^{\frac{\rho-2-n(\rho+1)i}{3n}}, \quad t>t_0,\quad i=1,\dots,4,
\end{gather}
where  $\gamma_i$, $i=0,\dots,4$, $\lambda$ and $\delta$ are arbitrary constants with $\gamma_0\lambda\delta\neq0$. Both equation and boundary conditions are invariant with respect to the scaling symmetry operator $Q=3nt\partial_t+(\rho+1)n x\partial_x+(\rho-2) u\partial_u$ (Case 1 of Table~1).
Using the corresponding ansatz (Case $1_{\rho\ne-1}$ of Table~4) this problem reduces to the initial value problem (IVP) for a fifth-order ODE,
\begin{gather}\arraycolsep=0ex
\begin{array}{l}
\delta\varphi'''''+ \lambda\varphi'''+\left(\varphi^n-\frac{\rho+1}{3}\omega\right)\varphi'+\frac{\rho-2}{3n}\varphi=0, \\[1ex]
\varphi(0)=\gamma_0, \quad
\dfrac{{\rm d}^{\,i} \varphi(\omega)}{{\rm d} \omega^i}\bigg|_{\omega=0}=\gamma_i,\quad i=1,\dots,4.
\end{array}
 \label{eq:IVP_from_BV_KdV}
\end{gather}
After the problem for the latter IVP  is solved numerically, then the corresponding solution of BVP~\eqref{BV_Kawahara}--\eqref{BV_Kawahara_bc} can be recovered using the similarity transformation $u=t^{\frac{\rho-2}{3n}}\varphi(\omega)$ with  $\omega=xt^{-\frac{\rho+1}{3}}$.

We illustrate the usage of Lie symmetries for the construction of numerical solutions  for the Kawahara equations with time-dependent coefficients by the following example.
\begin{example}
Consider the  equation
\begin{equation*}
v_t+v_x+\tfrac32\varepsilon v v_x+\tfrac12\varkappa v_{xxx}+\tfrac12\gamma v_{xxxxx}=0
\end{equation*}
that arises as a model describing the propagation of long nonlinear waves in the water covered by ice~\cite{Goncharenko&Yakovlev2005,Ilichev1989,Marchenko1988,Tkachenko&Yakovlev1999,Yakovlev&Goncharenko2010}.
Here
\[\varepsilon=\frac aH,\quad \varkappa=\frac{h}{\rho_{\omega}g\lambda^2}(\sigma_0-\sigma_{xx}),\quad \gamma=\frac{Eh^3}{12(1-\nu^2)\rho_{\omega}g\lambda^4},\]
where $v$ is the dimensionless amplitude of the oscillations of the under-ice surface
of the fluid about the horizontal equilibrium position, $a$ is the characteristic wave amplitude,
$H$ is the depth of the fluid, $2\pi\lambda$ is the characteristic wavelength, $\rho_{\omega}$ and $\rho_i$ are
the densities of the fluid and ice, respectively; $h$, $E$, and $\nu$ are the thickness, Young's modulus and
Poisson's ratio of the ice, and $\sigma_{xx}$ is a component of the ice sheet stress tensor, $\sigma_0=gH[\rho_{\omega}H/(3h)+\rho_i].$   It
is assumed that $\sigma_{xx}\approx10^5 {\rm N/m^2}$ is the result of external forces~\cite{Ilichev1989}.

\begin{figure}[b!]
\begin{minipage}[t]{75mm}
\centering
\includegraphics[height=60mm]{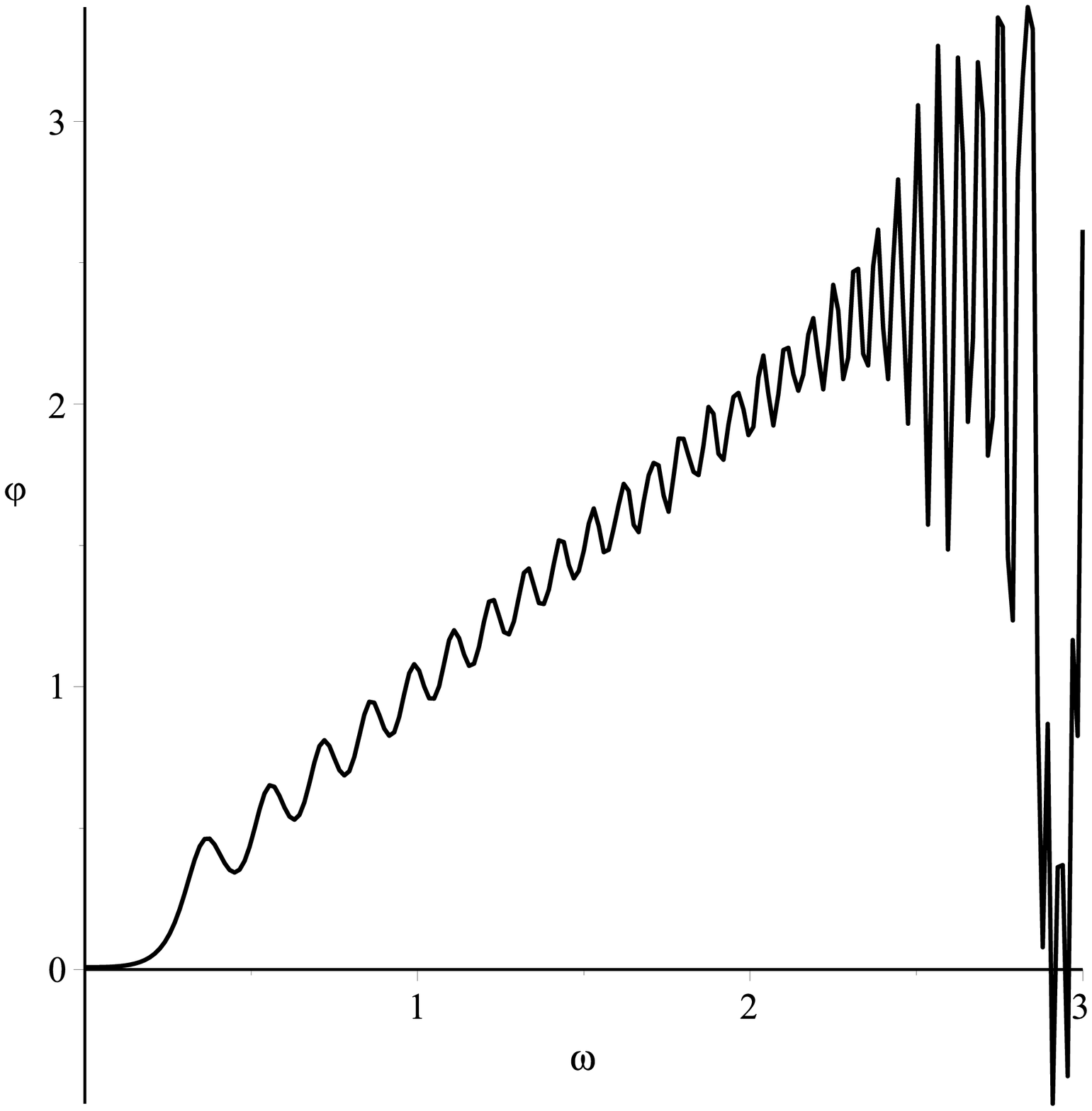}
\caption{\small Solution of IVP~\eqref{eq:IVP_from_BV_KdV1}, $\gamma_0=1/120$.}
\end{minipage}
\quad
\begin{minipage}[t]{75mm}
\centering
\includegraphics[width=72mm]{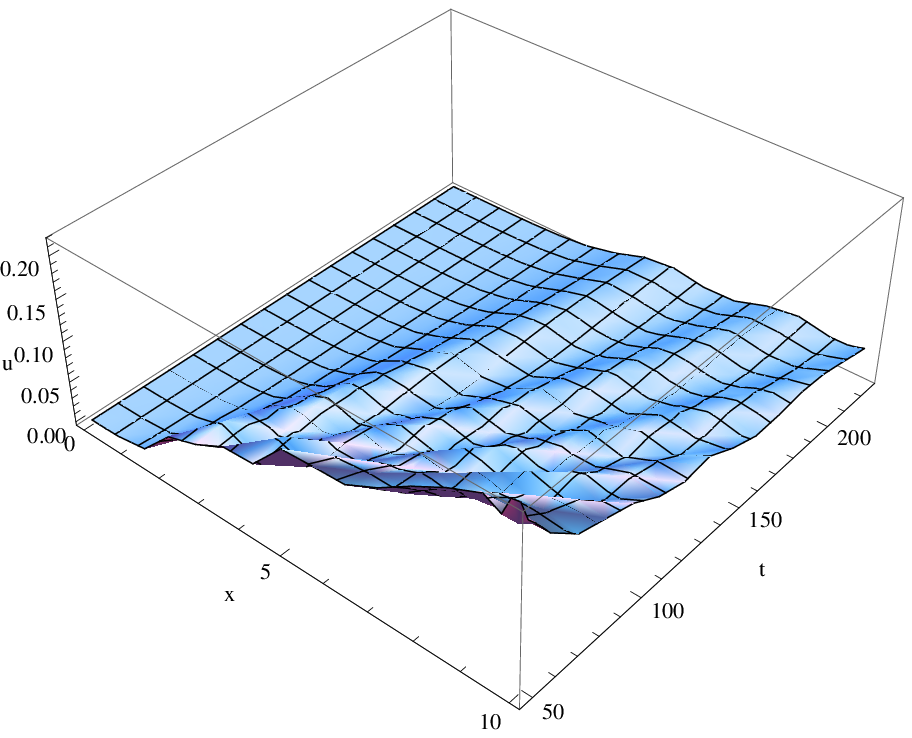}
\caption{\small Solution of BVP~\eqref{Kawahara_March}-\eqref{Kawahara_March_bc}, $\gamma_0=1/120$.
}
\end{minipage}
\end{figure}

We suppose that the growth of ice thickness is  described by the law $h=0.04\sqrt{t},$ which for certain weather conditions is in well agreement with the data  obtained for the sea of Azov for 10 days (240 hours) of observations of ice growth starting from h=0.1{\rm m}~\cite{Bukatov}.
Then for the values $\lambda\approx100{\rm m}$, $H\approx10{\rm m}$, $E\approx3\cdot10^9{\rm N/m^2}$, $a\approx 0.1{\rm m}$, $\rho_{\omega}\approx1030 {\rm kg/m^3}$, $\rho_{\omega}\approx 916{\rm kg/m^3}$ and $\sigma_0\approx1.2\cdot10^6{\rm N/m^2}$ that is calculated for average ice thickness $h_a\approx 0.3{\rm m}$ we will have a model equation of the form
\begin{equation*}
v_t+v_x+\alpha v v_x+\lambda t^\frac12 v_{xxx}+\delta t^\frac32 v_{xxxxx}=0,
\end{equation*}
where $\alpha=1.5\cdot10^{-2}$, $\beta\approx2.20215\cdot10^{-5}$ and $\delta\approx1.05566\cdot10^{-8}$ (after converting time in $E$, $\sigma_0$ and $\sigma_{xx}$ in hours). To reduce this equation to the form~\eqref{BV_Kawahara} we make the change of the dependent variable $u=1+\alpha v$ and get the equation
\begin{equation}\label{Kawahara_March}
u_t+uu_x+\lambda t^\frac12 u_{xxx}+\delta t^\frac32 u_{xxxxx}=0
\end{equation}
where $\lambda$ and $\delta$ remain the same.
We consider the  boundary conditions
\begin{equation}\label{Kawahara_March_bc}
u(t,0)=\gamma_0 t^{-\frac12},\  u_x(t,0)=0,\  u_{xx}(t,0)=0,\  u_{xxx}(t,0)=0,\  u_{xxxx}(t,0)=0,
\end{equation}
that are invariant with respect to  the operator of scaling symmetry $2t\partial_t+x\partial_x-u\partial_u$ of the latter equation. Such a BVP reduces to the following initial value problem
\begin{gather}\arraycolsep=0ex
\begin{array}{l}
\delta\varphi'''''+ \lambda\varphi'''+\left(\varphi-\frac{1}{2}\omega\right)\varphi'-\frac{1}{2}\varphi=0, \\[2ex]
\varphi(0)=\gamma_0, \quad \varphi'(0)=\varphi''(0)=\varphi'''(0)=\varphi''''(0)=0.
\end{array}
 \label{eq:IVP_from_BV_KdV1}
\end{gather}
The numerical solution for this initial value problem is presented on Fig.~4. The corresponding numerical solution of equation~\eqref{Kawahara_March}
with the associated boundary conditions~\eqref{Kawahara_March_bc} is presented on Fig.~5.
\end{example}

\section{Conclusion and discussion}

In the present paper the group classification problem for class~\eqref{eq_ggKawahara} of variable coefficient generalized Kawahara equations  was solved exhaustively. As a result, new variable coefficient nonlinear models admitting Lie symmetry extensions were derived.
This became possible due to an appropriate gauge of arbitrary elements of the class. Namely, the gauge $\alpha=1$ was utilized.
The use of different equivalence groups for the cases $n\neq1$ and $n=1$, which were found in the course of the study of admissible transformations in class~\eqref{eq_ggKawahara}, allowed us to write down the classification list in a simple and concise form (see Table~1). For convenience of further applications, in Table~2 we also presented the classification list extended by the equivalence transformations.
Then  one-dimensional subalgebras of Lie symmetry algebras admitted by equations from class~\eqref{eq_ggKawahara} were classified
and all inequivalent reductions  with respect to such subalgebras  were performed. It is obvious that the extension
of known solutions of constant coefficient equations from class~\eqref{eq_ggKawahara} by
 equivalence transformations of this class is a preferable way for the construction of exact solutions to the equations~\eqref{eq_ggKawahara} that are reducible to constant coefficient equations, i.e., to equations of the form~\eqref{eq_rKawahara1} and~\eqref{eq_rKawahara2}.
Some exact solutions for the classes of equations~\eqref{eq_rKawahara2} and equations~\eqref{eq_rKawahara1} with $n=2$ were constructed as illustrative examples. We also presented a class of boundary  value problems for variable coefficient Kawahara equations possessing scaling symmetry and constructed a numerical solution for the specific equation that could be of interest for applications.

In the framework of modern group analysis of differential equations the following problems for equations from class~\eqref{eq_ggKawahara} can also be studied.

$\diamond$ {\it The study of conservation laws.} Of course, using modern computer algebra packages it is easy to compute low-order conservation laws, not to mention zero-order conservation laws. The main problem is to prove that the set of orders of conservation laws is bounded and then to describe exhaustively the entire space of conservation laws. The obvious zero-order conservation laws of equations of the form~\eqref{eq_gKawahara}
 are given by conserved vectors with characteristics $1$ and $u$
\begin{gather*}
\left(u,\tfrac1{n+1}\,{\alpha(t)}u^{n+1}+\beta(t)u_{xx}+\sigma(t)u_{xxxx}\right),\\[1ex]
\left(\tfrac12 u^2, \tfrac1{n+2}\,{\alpha(t)}u^{n+2}+\beta(t)\left(uu_{xx}-\tfrac12u_x^2\right)+\sigma(t)\left(uu_{xxxx}-u_xu_{xxx}+\tfrac12u^2_{xx}\right)\right).
\end{gather*}
These are conservation laws of  momentum and energy, respectively.

$\diamond$ {\it Potential symmetries}~\cite{Bluman&Reid&Kumei1988,Pucci&Saccomandi1993}. For each equation of the form~\eqref{eq_ggKawahara}
 we can construct the potential systems,
\begin{gather*}
v_x=u,\quad v_t=-\tfrac1{n+1}\,{\alpha(t)}u^{n+1}-\beta(t)u_{xx}-\sigma(t)u_{xxxx},\\[1ex]
v_x=\tfrac12u^2,\quad v_t=-\tfrac1{n+2}\,{\alpha(t)}u^{n+2}-\beta(t)\left(uu_{xx}-\tfrac12u_x^2\right)-\sigma(t)\left(uu_{xxxx}-u_xu_{xxx}+\tfrac12u^2_{xx}\right),
\end{gather*}
associated with the above conservation laws.
The complete description of  potential systems needs the exhaustive classification of local conservation laws of such equations~\cite{Bluman&Cheviakov&Anco2010,ipsv,Popovych&Kunzinger&Ivanova2008}. Then Lie symmetries of the constructed potential systems should be found, which may result in nontrivial potential symmetries for the generalized Kawahara equations.

In these problems the investigation can be restricted to subclass~\eqref{eq_gKawahara} of class~\eqref{eq_ggKawahara} without loss of generality, i.e., it can be assumed that $\alpha=1$. The justification is presented in Section~2.

\subsection*{Acknowledgements}
The authors wish to thank members of Department of Applied Research of Institute of Mathematics of NASU and especially R.O. Popovych for useful discussions and also unknown referees for the suggestions that have led to improvement of the paper. O.V. expresses the gratitude to the hospitality
shown by the Czech Technical University in Prague during her visit to the University.

\end{document}